\documentclass[aps,twocolumn,prb,showpacs]{revtex4}
\usepackage{graphicx}
\usepackage{color}
\usepackage{amsmath}
\usepackage[bf,Large]{}

\begin{document}

\title{Induced superconductivity in noncuprate layers of the
  Bi$_2$Sr$_2$CaCu$_2$O$_{8+\delta}$ high-temperature superconductor:
  Modeling of scanning tunneling spectra}

\author{Ilpo Suominen}
\affiliation{Department of
Physics, Tampere University of Technology, P.O. Box 692, FIN-33101
Tampere, Finland}

\author{Jouko Nieminen}
\email{jouko.nieminen@tut.fi}
\affiliation{Department of
Physics, Tampere University of Technology, P.O. Box 692, FIN-33101
Tampere, Finland}
\affiliation{Department of Physics, Northeastern
University, Boston, MA 02115, USA}

\author{R.S. Markiewicz} \author{A. Bansil}
\affiliation{Department of Physics,
Northeastern University, Boston, MA 02115, USA}

\date{Version of \today}

\begin{abstract}
We analyze how the coherence peaks observed in Scanning Tunneling 
Spectroscopy (STS) of cuprate high temperature superconductors are 
transferred from the cuprate layer to the oxide layers adjacent to the 
STS microscope tip. For this purpose, we have carried out a realistic 
multiband calculation for the superconducting state of 
Bi$_2$Sr$_2$CaCu$_2$O$_{8+\delta}$ (Bi2212) assuming a short range d-wave 
pairing interaction confined to the nearest-neighbor Cu $d_{x^2-y^2}$ 
orbitals. The resulting anomalous matrix elements of the Green's function 
allow us to monitor how pairing is then induced not only within the 
cuprate bilayer but also within and across other layers and sites. The 
symmetry properties of the various anomalous matrix elements and the 
related selection rules are delineated.
\end{abstract}

\date{Version of \today}

\maketitle

\section{Introduction}

Scanning tunneling spectra (STS) of the cuprates 
\cite{Fischer,McElroy,Hudson,Pan,Yazdani} clearly show the presence of 
superconducting gaps and the associated coherence peaks.  The `leaking' of 
superconductivity from the cuprate layers into the oxide layers is a form 
of proximity effect \cite{meissner, degennes, McMillan}. A recent STS 
study\cite{parker} finds that the magnitude of the superconducting gap or 
the pseudogap is not solely determined by the local doping, but is also 
sensitive to the nearby nanoscale surroundings, raising the broader 
question as to how superconductivity is transfered across various 
orbitals/sites in the cuprates.\cite{hoffman} In this connection, we have 
recently developed a Green's function based methodology for carrying out 
realistic computation of scanning tunneling microscopy/spectroscopy 
(STM/STS) spectra in the normal as well as the superconducting state of 
complex materials, where the nature of the tunneling process, i.e., the 
effect of the tunneling matrix element is properly taken into account. In 
our approach, all relevant orbitals in the material are included in a 
multi-band framework, and the tunneling current is computed directly for a 
specific tip position on the semi-infinite surface of the solid. An 
application to the case of overdoped Bi2212 was reported in Refs. 
\onlinecite{NLMB} and \onlinecite{nieminenPRB}, where it was shown, for 
example, that the striking asymmetry of the STS spectrum between high 
positive and negative bias voltages arises from the way electronic states 
in the cuprate layer couple to the tip: With increasing negative bias 
voltage, new tunneling channels associated with $d_{z^2}$ and other 
orbitals begin to open up to yield the large tunneling current. The 
asymmetry of the tunneling current at high energies could thus be 
understood naturally within the conventional picture, without the need to 
invoke exotic mechanisms. Results of Refs. \onlinecite{NLMB} and 
\onlinecite{nieminenPRB} show clearly that the STS spectrum is modified 
strongly by matrix element effects as has been shown 
previously for angle-resolved photoemission\cite{photo}, resonant 
inelastic x-ray scattering\cite{rixs}, and other highly resolved 
spectroscopies.\cite{compton,magcompton,positron}.

The STM/STS modeling in Refs. \onlinecite{NLMB} and
\onlinecite{nieminenPRB} is based on invoking the common assumption
that the pairing interaction in cuprates is d-wave, involving nearest
neighbor $d_{x^2-y^2}$ orbitals of Cu atoms.
Nevertheless, our computed STS spectrum reproduces, in accord with
experimental observations, the superconducting gap and coherence peaks
at the position of the tip, even though the tip is not in direct contact
with the cuprate layer. Our STS modeling scheme thus provides a
natural basis for examining how the pairing interaction, which is
limited to nearest-neighbor Cu $d_{x^2-y^2}$ orbitals in our
underlying Hamiltonian, gets transferred to other layers and sites.

This article attempts to address these and related issues with the example 
of overdoped Bi2212. Central to our analysis is the concept of tunneling 
channels, which allows us to identify the contribution to the total 
tunneling current from individual sites/orbitals in the semi-infinite 
solid. Moreover, we can distinguish between regular and anomalous 
contributions to the tunneling signal, which arise from the corresponding 
matrix elements of the Nambu-Gorkov Green's function tensor. The anomalous 
channels are physically related to the formation and breaking up of Cooper 
pairs. In particular, matrix elements of the anomalous Green's function 
can be used to monitor the contribution to the coherence peaks in the STS 
spectrum resulting from specific orbitals/sites in the material. In this 
way, we delineate how the pairing amplitude travels from the 
nearest-neighbor Cu-sites to other sites and orbitals within the cuprate 
plane as well as outside to the second cuprate plane and to the BiO/SrO 
layers. The symmetry properties of various matrix elements are analyzed 
and related selection rules are worked out. 

An outline of this article is as follows. The introductory remarks are
followed in Section II with an overview of the relevant methodological
details of the underlying Hamiltonian and of our STS formalism. Section
III discusses proximity effects and is divided into several
subsections, which address pairing amplitudes in various
layers. Section IV discusses selection rules and issues related to the
symmetry of the gap through an analysis of the anomalous matrix
elements. It is divided into consideration of on-site cases where the
pairing orbitals lie at the same horizontal position, and to cases where
these orbitals lie at other sites in the lattice. Finally, Section
V presents a concluding discussion and a summary of our
results. The added Appendix clarifies the symmetry properties of
the regular matrix elements, which play an important role in the
analyis of the symmetry of the anomalous matrix elements of the
Green's function.

\begin{figure}
    \includegraphics[width=0.5\textwidth]{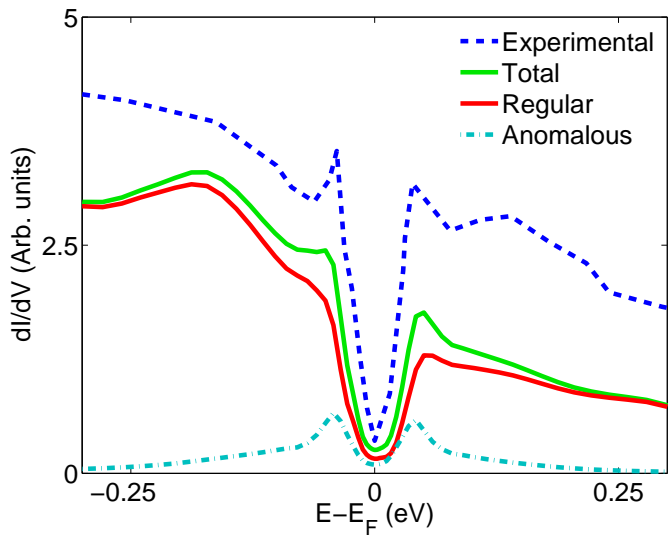}
    \caption{(Color online)
Theoretical (green) STS spectrum normalized as shown 
in the figure is compared with the experimental (dashed blue) spectrum 
\cite{McElroy} in optimally doped Bi2212. Regular (red) 
and anomalous (turquoise) parts of the computed spectrum are shown 
separately. All computations are based on 
      Eq. \eqref{spectralfunction}.
Coherence peaks arise only from the anomalous component of the Green's 
function.
}
\label{fig0ab}
\end{figure}

\section{Description of the model}

The model underlying our analysis is the same as in Ref.
\onlinecite{nieminenPRB} to which we refer for details. An overview is
nevertheless presented for completeness, and to introduce the various
quantities needed for the present study. The Bi2212 sample is modeled
as a slab of seven layers terminated by the BiO layer, which is
followed by layers of SrO, CuO$_2$, Ca, CuO$_2$, SrO, and
BiO.\cite{ABfoot1,Spathis,Hirschfeld} The tunneling current is
computed using a $2\sqrt{2} \times 2\sqrt{2}$ real space supercell
consisting of 8 primitive surface cells with a total of 120 atoms. The
crystal structure is taken from Ref. \onlinecite{Bellini}.  The STM
tip is modeled as an s-orbital lying at the apex of the tip. The
electron and hole orbitals included in the computations are:
($s,p_x,p_y,p_z$) for Bi, Ca and O; $s$ for Sr; and
($4s,d_{3z^2-r^2},d_{xy},d_{xz},d_{yz}, d_{x^2-y^2}$) for Cu
atoms. This yields $2 \times 58$ electron (spin up) and hole (spin
down) orbitals in the primitive unit cell, or a total of $2 \times
464$ orbitals in the simulation supercell. The Green's function is
computed using $256$ equally distributed {\bf k}-points in the
supercell 
which corresponds to $8 \times 256 = 2048$ {\bf k}-points in a
primitive cell.

The multiband Hamiltonian in which superconductivity is included by adding 
a pairing interaction term $\Delta$ is\cite{ABfoot4,Das,Basak}
\begin{equation}
  \begin{array}{ccc}
\hat{H}& = & \sum_{\alpha\beta\sigma}
\left[\varepsilon_{\alpha}c^{\dagger}_{\alpha \sigma} c_{\alpha \sigma}+
V_{\alpha \beta}
c^{\dagger}_{\alpha \sigma} c_{\beta\sigma}\right]\\
&+&
\sum_{\alpha \beta
\sigma} \left[\Delta_{\alpha \beta} c^{\dagger}_{\alpha \sigma}
c^{\dagger}_{\beta -\sigma} + \Delta_{\beta \alpha}^{\dagger}
 c_{\beta -\sigma} c_{\alpha \sigma} \right]
  \end{array}
\label{hamiltonian}
\end{equation}
with real-space creation (annihilation) operators $c^{\dagger}_{\alpha
  \sigma}$ (or $c_{\alpha \sigma}$). Here $\alpha$ is a composite
index identifying both the type of orbital and its site, and $\sigma$
is the spin index.  $\varepsilon_\alpha$ denotes the on-site energy of
the $\alpha^{th}$ orbital, and $V_{\alpha\beta}$ is the hopping
integral between the $\alpha$ and $\beta$ orbitals.  The hopping
parameters are chosen to reproduce the LDA
bands.\cite{RMfoot1,ABfoot2,AB1,AB2,AB3}

In the mean field approximation, the coupling between electrons and holes
is of the form
\begin{equation}
  \label{eq:gapgen}
  \Delta_{\alpha \beta} = \sum_{a b} U_{\alpha  \beta a b} \langle c_{a
  \downarrow}c_{b \uparrow} \rangle. 
\end{equation}
Since the interaction $U$ is not known, the standard practice is to 
introduce a gap parameter, which gives the correct gap width and 
symmetry\cite{Flatte}. Specifically, we take $\Delta$ to be non-zero 
only between the $d_{x^2 - y^2}$ orbitals of the nearest neighbor Cu 
atoms, and to possess a d-wave form, i.e., $\Delta_{d (d \pm x)} = +\vert 
\Delta \vert$ and $\Delta_{d (d \pm y)} = -\vert \Delta \vert$, where $d$ 
denotes the $d_{x^2-y^2}$ orbital at a chosen site, and $d \pm x/y$ the 
$d_{x^2-y^2}$ orbital of the neighboring Cu atom in 
$x/y$-direction.\cite{ABfoot3} This 
form allows electrons of opposite spins to combine to produce 
superconducting pairs such that the resulting superconducting gap is zero 
along the nodal directions $k_x=\pm k_y$, and is maximum along the 
antinodal directions. The gap parameter value of $\vert\Delta\vert = 45 
meV$ is chosen to model a typical experimental spectrum\cite{McElroy} for 
our illustrative purposes.\cite{RMfoot2,Fang}

We discuss pairing between different orbitals in terms of the 
tensor (Nambu-Gorkov) Green's function ${\cal G}$ (see, e.g.,
Ref. \onlinecite{Fetter}) 
\begin{displaymath}
  {\cal G} =
\left(
   \begin{array}{cc}
G_{e}& F\\
F^{\dagger}& G_{h}
   \end{array}
\right)
\end{displaymath}
where $G_{e}$ and $G_{h}$ denote the electron and hole Green's function, 
respectively.

The following expressions for the pairing amplitudes in a tight-binding 
basis, which are derived in Ref. \onlinecite{nieminenPRB}, are 
especially relevant for our analysis. 
\begin{equation}
\langle c_{\alpha \downarrow}c_{\beta \uparrow} \rangle = \int
d\varepsilon [1-2f(\varepsilon)] \rho^{eh}_{\alpha \beta}(\varepsilon),
\label{ccF}
\end{equation}
where the density matrix is
\begin{displaymath}
 \rho^{eh}_{\alpha \beta}(\varepsilon) = -\frac{1}{\pi}Im[F^{+}_{\alpha
 \beta}(\varepsilon)]. 
\end{displaymath}
Here,
$ F^{+}_{\alpha \beta}(\varepsilon)$ can be solved by using the
tensor form of Dyson's equation for the retarded Green's 
function. Similarly, 
\begin{equation}
\langle c^{\dagger}_{\alpha \uparrow}c^{\dagger}_{\beta \downarrow}
\rangle = \int d\varepsilon [1-2f(\varepsilon)] \rho^{eh\dagger}_{\beta
\alpha}(\varepsilon).
\label{ccF1}
\end{equation}

Eqs. \eqref{ccF} and \eqref{ccF1} reveal the relationship between the 
anomalous part of the Green's function tensor and the pairing amplitudes 
between various sites. In particular, symmetry properties of $F_{\alpha 
\beta}$ are seen to be related directly to those of $\langle c_{\alpha
  \downarrow}c_{\beta \uparrow} \rangle$.


The tunneling spectrum is computed by using the Todorov-Pendry 
expression \cite{Todorov,Pendry} for the
differential conductance $\sigma$ between orbitals of the tip ($t,t'$)
and the sample ($s,s'$), which in our case can be written as
\begin{equation}
\sigma = \frac{dI}{dV} = \frac{2 \pi e^2 }{\hbar} \sum_{t t' s s'}
\rho_{tt'}(E_F)V_{t's} \rho_{ss'}^{}(E_F+eV)V_{s't}^{\dagger}.
\label{conductance}
\end{equation}

Since electrons are not eigenparticles in the presence of the pairing
term, 
the density matrix can be rewritten by applying the tensor form of
Dyson equation \cite{nieminenPRB}:
\begin{equation}
\rho_{s s'} =
 -\frac{1}{\pi}\sum_{\alpha} (G_{s \alpha}^{+} \Sigma{''}_{\alpha} G_{\alpha
s'}^{-} + F_{s \alpha}^{+} \Sigma{''}_{\alpha} F_{\alpha
s'}^{-}),
\label{spectralfunction}
\end{equation}
where $\Sigma{''}_{\alpha}$ is the imaginary part of 
self-energy.\cite{ABfoot5,elphon} The left-hand side of Eq. 
\eqref{spectralfunction} is the ordinary density matrix for electrons, 
which is equivalent to the traditional Tersoff-Hamann approach 
\cite{Tersoff-Hamann}. However, as discussed in Ref. 
\onlinecite{nieminenPRB}, our decomposition of the spectrum into tunneling 
channels in Eq. \eqref{spectralfunction} provides a powerful way to gain 
insight into the nature of the STS spectrum, especially in complex 
materials.\cite{Sautet, Magoga} Note that the right side of Eq. (6) 
contains terms originating from the anomalous part of the Green's 
function. In Ref.  \onlinecite{nieminenPRB} we showed that coherence peaks 
appear only through the matrix elements of the anomalous Green's function. 
This role of the anomalous terms is demonstrated in Fig. ~\ref{fig0ab}, 
where we see that the coherence peaks are absent in the partial spectrum 
resulting from the regular terms of the Green's function (red curve).

\begin{figure*}[th]
    \includegraphics[width=1.00\textwidth]{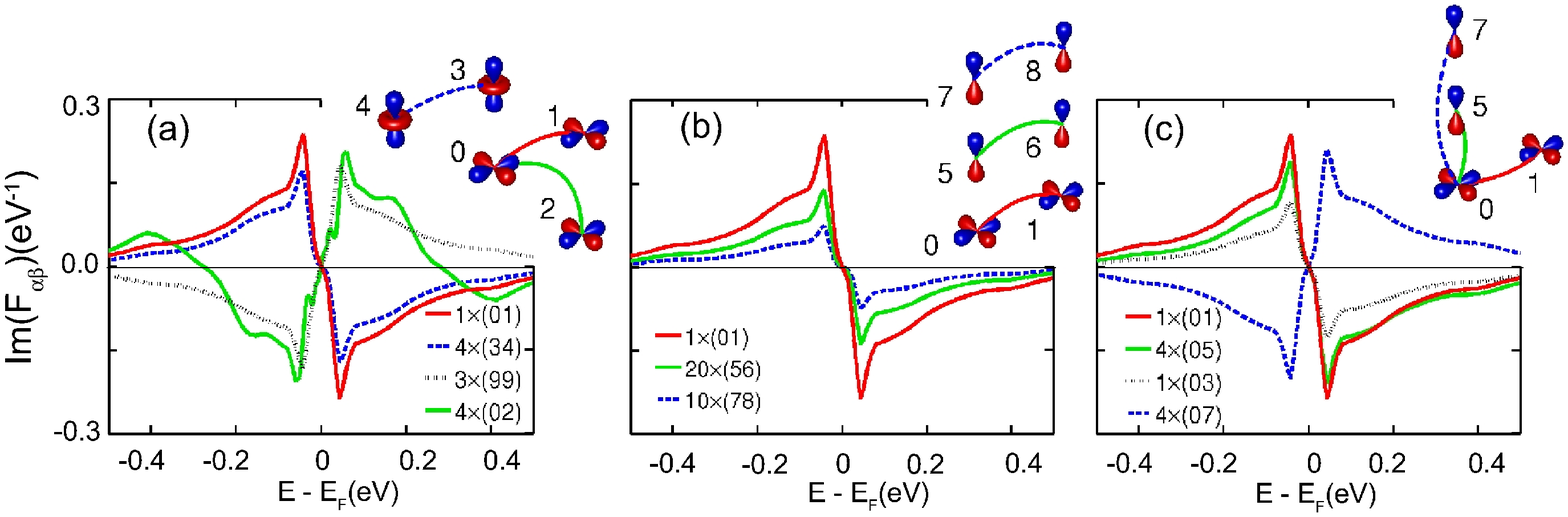}
  \caption{(Color online) 
    Imaginary part of the matrix elements of the anomalous Green's
    function $F_{\alpha\beta}$ for various ($\alpha \beta$) pairs.
    [Recall that $\alpha$ and $\beta$ are composite indices denoting
    both site and orbital.] Meaning of values of $\alpha$ and $\beta$,
    which range from 0 to 9, is explained in Table 1. For example,
    index 0 refers to the $d_{x^{2}-y^{2}}$ orbital on the central Cu
    atom in the cuprate plane nearest to the STM tip, and index 1 to
    the $d_{x^{2}-y^{2}}$ orbital on nearest neighbor Cu atom in the
    same cuprate plane. The main matrix element between the two
    preceding orbitals, i.e., the (01) element, is shown by red lines
    for reference in all panels. Other matrix elements are shown
    scaled by factors ranging from 2-20 as indicated in the legends.
    Symmetries of the orbitals involved in various cases are shown
    schematically in the upper right hand side portions of the
    figures.  Matrix elements compared with (01)(red) are: (a) (02)
    (green), (34) (dashed blue) and (99) (dotted black) for pairing
    within a CuO$_2$ bilayer; (b) (56) (green) and (78) (dashed blue)
    for intralayer pairing in SrO and BiO layers; and (c) (03) (dotted
    black), (05) (green) and (07) (dashed blue) for pairing of CuO$_2$
    along the line connecting the central Cu and the surface Bi.
}
  \label{fig1ab}
\end{figure*}

\section{Interlayer and intralayer proximity effects}

In this section, we analyze the induced pairing amplitude $\langle
c_{\alpha \downarrow}c_{\beta \uparrow} \rangle$ for a representative
set of orbitals. It will be seen that despite the short range of the
pairing interaction $\Delta_{\alpha \beta}$, the anomalous Green's
function, $F_{\alpha \beta}$, possesses a longer range. More
specifically, we delineate induced pairing effects as follows: (i)
Within a CuO$_2$ bilayer (Fig. \ref{fig1ab}(a)); (ii) Intra-layer
pairing in SrO and BiO layers (Fig. \ref{fig1ab}(b)); and (iii)
Interlayer pairing between CuO$_2$ and SrO/BiO layers
(Fig. \ref{fig1ab}(c)). We discuss each of these cases in turn below
with reference to Fig. \ref{fig1ab} and Table \ref{tab:orbitals}.

\begin{table}[htbp]
  \centering
  \begin{tabular}{|c|c|c|c|}
\hline
label&orbital&atom&layer\\
\hline
0&$d_{x^2-y^2}$&Cu (c)&CuO$_{2}$ (1st)\\
1&$d_{x^2-y^2}$&Cu (nn)&CuO$_{2}$ (2nd)\\
2&$d_{x^2-y^2}$&Cu (nn)&CuO$_{2}$ (2nd)\\
\hline
3&$d_{z^2}$&Cu (c)&CuO$_{2}$ (1st)\\
4&$d_{z^2}$&Cu (nn)&CuO$_{2}$ (1st)\\
\hline
5&$p_{z}$&O (c)&SrO \\
6&$p_{z}$&O (nn)&SrO\\
\hline
7&$p_{z}$&Bi (c)&BiO \\
8&$p_{z}$&Bi (nn)&BiO\\
\hline
9&$p_{x}$&O (b)&CuO$_{2}$ (1st)\\
\hline
  \end{tabular}
  \caption{
    Shorthand notation used for the indices $\alpha$ and $\beta$ in 
    Fig. \ref{fig1ab} is 
    defined. For each of the indices, varying from 0 to 9, 
    the table gives the atomic site [central (c), nearest neighbor (nn) 
    and bonding (b)], the orbital and the layer involved. Order 
    of the layers is: BiO, SrO, CuO$_2$ (1st) and CuO$_2$ (2nd), where 
    BiO is the termination layer which lies closest to the STM tip.
  }
  \label{tab:orbitals}
\end{table}

\subsection{Pairing within CuO$_2$ bilayer}

The most important anomalous matrix elements within the CuO$_2$ bilayer 
are shown in Fig. \ref{fig1ab}(a). The red curve gives the contribution 
$F_{01}$ from $d_{x^2-y^2}$ orbitals of two neighbouring Cu atoms in the 
$x$-direction, i.e., the matrix element between a spin up electron orbital 
at a Cu-site and a spin down hole orbital at the neighbouring Cu-site. 
This is the principal pairing matrix element since in our model 
$\Delta_{\alpha \beta}$ is non-zero only between two such orbitals. 

The matrix element between the $d_{x^2-y^2}$ orbitals of two Cu atoms
at the same horizontal position within the CuO$_2$ bilayer is zero by 
symmetry.
However, the matrix element $F_{02}$ between a Cu atom in the upper
layer and each of the four neighbouring Cu atoms of the lower layer
(and vice versa) is seen from Fig.\ref{fig1ab}(a) to be substantial
with an amplitude which is about $1/4$th of $F_{01}$. This result shows
that pairing is not restricted to $d_{x^2-y^2}$ orbitals within a
single CuO$_2$ layer, i.e. it is not two-dimensional but extends
vertically within the bilayer.

The $d_{z^2}$ orbital of Cu also plays an important role. In fact, this 
orbital serves as a kind of gate for passing tunneling current from the 
cuprate layers to the SrO and BiO layers. Fig. \ref{fig1ab}(a) shows that 
the amplitude $F_{34}$ is about 1/5th of $F_{01}$ and comparable to 
$F_{02}$. There is also a smaller (about 1/5th of $F_{34}$) rotationally 
invariant matrix element $F_{04}$ (not shown in Fig. \ref{fig1ab}) between the 
$d_{z^2}$ of a central Cu and the $d_{x^2-y^2}$ orbitals of the four 
neighbouring Cu atoms. At first sight this seems to break the d-wave 
symmetry, but we will show below that the combined symmetry of the 
orbitals involved remains d-wave \cite{rotationfoot}.

The role of O-atoms in the cuprate layers can be delineated through
the matrix elements $F_{99}$ and $F_{09}$. In Fig.  \ref{fig1ab}(a) we
show the onsite matrix element $F_{99}$, which is about 1/4th of the
$F_{01}$ term.  We observe that in real space rotations of $\pi/2$
around the central Cu, $F_{99}$ changes it sign. A smaller
contribution is found for $F_{09}$ (not shown in
Fig. \ref{fig1ab}). Its symmetry properties are consistent with the
symmetry of the Zhang-Rice singlet, where a local orbital is
constructed as a linear combination of the four oxygen atoms around
the central Cu.  The symmetry analysis of Section
\ref{sec:symmetryrules} below shows that both $F_{99}$ and $F_{09}$ are
also consistent with the d-wave symmetry.

Finally, we note that there is a substantial term, $F_{03}$, 
between the $d_{z^2}$ and $d_{x^2-y^2}$ orbitals
of {\it the same Cu atom}, which is 
perhaps surprising. Fig. \ref{fig1ab}(c) shows that this pairing 
amplitude is about half of $F_{01}$. Since this is an onsite term, the 
d-wave symmetry again follows from the combined symmetry of the two 
orbitals, as discussed in Section \ref{sec:symmetryrules} 
below.

\subsection{Intralayer pairing within SrO and BiO layers }

In considering intralayer pairing in the SrO/BiO layers, we find that
for Bi or apical O atoms, the most important non-zero anomalous matrix
elements occur between $p_z$-orbitals of the central atom and its four
neighbours, i.e. $F_{56}$ and $F_{78}$. These matrix elements possess
the same d-wave symmetry as $F_{01}.$ While all matrix elements have
the same energy dependence in Fig. \ref{fig1ab}(b), $F_{01}$ is about
30 times larger than $F_{56}$ or $F_{78}.$ The coherence peaks lie at
exactly the same energy in each layer, i.e., the gap width is the same
in all layers. The scaling factor for the amplitude seems to roughly
follow the spectral weight of the orbitals. Hence, the pairing of
electrons within the oxide layers seems to be a direct consequence of
the tail of the CuO$_2$ electron wave function within the various
layers.  This kind of pairing is in the spirit of the original idea of
proximity effect \cite{meissner, degennes, McMillan} where superconductivity is
viewed as ``leaking'' from the superconducting part of the sample to
the normal state material. Although the aformentioned orbitals are the
most important ones, non-zero pairing is not restricted to just these
orbitals. On the other hand, certain terms are strictly zero due to
symmetry. In particular, all the onsite F$_{ss}$, F$_{p_{x}p_{x}}$,
F$_{p_{y}p_{y}}$, and F$_{p_{z}p_{z}}$ from Bi and O(Sr) are zero, as
are many Bi-O(Bi) and O(Sr)-Sr terms. However, $F_{p_{x}p_{y}}$ of two
neighbouring Bi's is non-zero as is $F_{p_{x}p_{y}}$ on the same Bi
atom.

\subsection{Interlayer pairing between CuO$_2$ and SrO/BiO layers }

Pairing on BiO and SrO layers is not restricted to intralayer terms 
discussed above. The interlayer terms $F_{05}$ and $F_{07}$ between the 
$p_z$-orbitals of the central Bi [or O(Sr)] and $d_{x^2-y^2}$ of Cu {\it 
right below} these atoms is, in fact significant, while the anomalous term 
to the neighbouring Cu atoms is rather small. The existence of these 
matrix elements might be surprising, since the regular matrix elements are 
zero by symmetry between $d_{x^2-y^2}$ and the rotationally symmetric 
orbitals of the atoms above the central Cu (See Appendix 
\ref{app:symmetry}). But we will show in the following section that these 
matrix elements are not symmetry forbidden. From Fig. \ref{fig1ab}(c), the 
scaling 
factor between these elements and $F_{01}$ is of the order of 4. Notably, 
this interlayer pairing would appear as $k_z$ dependence in the 
gap-function. If we make a reflection of the slab with respect to the Ca 
plane lying between the two CuO$_2$ layers, the corresponding anomalous 
matrix elements change their sign, indicating that this term has a node at 
$k_z =0$, and thus deviations from d-wave symmetry should be found for 
non-zero $k_z.$ The $k_{z}$-dependence is also seen in the rather small 
terms between $p_{z}$ orbitals of the Bi atoms of the surface layer, and 
the nearest neighbour Bi atoms of the BiO layer half the primitive cell 
below the surface.

\section{Symmetry and selection rules for induced pairing}
\label{sec:symmetryrules}

We now discuss the symmetry properties and the related selection rules for 
the anomalous matrix elements of the Green's function in terms of the 
d-wave symmetry of the pairing matrix.

\subsection{Symmetry properties of the anomalous matrix elements}

Note first that the symmetry of the pair wave function depends on the 
relative motion of the pairing electrons, i. e., only on the relative 
coordinate $R_i-R_j$. The analysis of the symmetry properties however 
becomes more transparent in $k$-space. Accordingly, we transform the real 
space matrix elements $F_{i\alpha j \beta}(\varepsilon)$ into k-space as 
\begin{equation}
  \label{eq:fourierF}
\begin{split}
  F_{\alpha \beta}(\mathbf{k},\varepsilon) =& \sum_{j}\langle \mathbf{k} \vert 0
  \alpha \rangle F_{0\alpha j \beta}(\varepsilon) \langle j \beta \vert
  \mathbf{k} \rangle \\ = & \sum_{j} F_{0 \alpha j \beta}(\varepsilon) e^{-i
    \mathbf{k} \cdot \mathbf{R_{j}}}
   \varphi^{*}_{\alpha}(\mathbf{k}) \varphi_{\beta}(\mathbf{k}).
\end{split}
\end{equation}
Here, we have set $R_i=0$ and $ \varphi_{\alpha}(\mathbf{k})$ is the 
orbital wave function in $k$-space. The site indices $i$ and $j$ and the 
orbital indices $\alpha$ and $\beta$ are shown explicitly for all matrix 
elements. The summation is taken over the site index $j$. The orbital 
indices are obviously not involved in the transformation. For simplicity, 
we will restrict the analysis below such that $j$ is either on-site or one 
of the nearest neighbors of the central site. The generalization to 
farther out neighbors is straightforward.

We need to take into account not only the phase difference between the 
sites, but also the form of the tight binding orbitals $ 
\varphi_{\alpha}(\mathbf{k}).$ These orbitals have the same symmetry in 
real-space and $k$-space. In particular,
\begin{equation}
\label{korbi}
\begin{split}
 \varphi^{*}_{x}(\mathbf{k}) =\langle {\mathbf{k}} \vert p_x \rangle
  \propto {k_x \over k}\\
 \varphi^{*}_{y}(\mathbf{k}) =\langle {\mathbf{k}} \vert p_y \rangle
 \propto {k_y \over k}\\
 \varphi^{*}_{x^2-y^2}(\mathbf{k}) =\langle {\mathbf{k}} \vert d_{x^2-y^2} \rangle
\propto
{k^2_{x} - k^2_{y} \over k^2}\\
 \varphi^{*}_{3z^2-r^2}(\mathbf{k}) =\langle {\mathbf{k}} \vert d_{3z^2-r^2} \rangle
\propto  {3k^2_{z} - k^2 \over k^2}.
\end{split}
\end{equation}

The symmetry of the matrix elements is now readily analyzed. We give two 
examples to illustrate the procedure: (i) $F_{01}$ between the 
$d_{x^2-y^2}$ orbitals of the central Cu and its neighbours; and (ii) 
$F_{03}$ between $d_{z^2}$ and $d_{x^2-y^2}$ at the central site. In the 
case of $F_{01}$ the product of the orbital functions is even under 
rotations by $\pi/2$:
\begin{displaymath}
  \varphi^{*}_{\alpha}(\mathbf{k}) \varphi_{\alpha}(\mathbf{k}) = \vert  \varphi_{\alpha}(\mathbf{k}) \vert^2  
\end{displaymath}
In fact, this applies to all cases where $\alpha = \beta.$ Summing over 
the four sites around the central Cu and applying the odd parity with 
respect to $\pi/2$ rotation of the real space matrix elements $F_{0
  \alpha j \alpha}\sim \Delta_{0j}$, we obtain
\begin{equation}
  \label{eq:x2y2symm}
  F_{\alpha \alpha}(\mathbf{k},\varepsilon) = 2 \vert F_{0 \alpha j\alpha}(\varepsilon) \vert
  \left( \cos{(k_x a)} - \cos{(k_y a)} \right)  \vert  \varphi_{\alpha}(\mathbf{k}) \vert^2,
\end{equation}
which is obviously d-wave. This is easy to see for the $d_{x^2-y^2}$ 
orbitals, but Eq. \eqref{eq:x2y2symm} leads to the same conclusion for any 
set of four neighbouring orbitals similar to the central one, as long as 
the real space element is odd under rotations by $\pi/2.$

Turning to the case of $F_{03}$, here the sum in Eq. \eqref{eq:fourierF}
consists of a single term (central site), so that there is no site
related phase factor.  We only need to consider the product of
orbitals:
\begin{equation}
  \label{eq:z2x2y2}
   \varphi^{*}_{3z^2-r^2}(\mathbf{k}) \varphi_{x^2-y^2}(\mathbf{k}) \propto
\frac{3k_{z}^2-k^2}{k^4}\left(k_{x}^2-k_{y}^2\right)
\end{equation}
This is the only term through which angular dependence enters in Eq. 
\eqref{eq:fourierF}. This again is d-wave, keeping in mind that only the 
in-plane symmetry is relevant. These considerations apply more generally 
to any case where the two orbitals involved lie at the same horizontal 
position with one of the orbitals being rotationally invariant and the 
other is d-wave.

\subsection{Selection rules for anomalous matrix elements}

Selection rules for the matrix elements $F_{i\alpha j\beta}$ of the 
anomalous Green's function do not follow directly from those for the 
corresponding regular matrix elements discussed in Appendix A. For this 
purpose, we write $F_{i\alpha j\beta}$ as\cite{nieminenPRB} 
\begin{equation}
F_{i\alpha j\beta}(\varepsilon)
= -G^{+}_{i \alpha k \gamma}(\varepsilon) \Delta_{k \gamma l \delta} G^{0-}_{l \delta j \beta}(-\varepsilon)
  \label{route_equation}
\end{equation}
where the Einstein summation convention is implicit, and both the Green's 
functions on the right hand side of the equation are regular. The first is 
the renormalized Green's function for the superconducting state, while the 
second with superscript zero is the bare Green's function for the normal 
state. However, as shown in Appendix \ref{app:symmetry}, the symmetry 
properties of these two Green's functions are the same since both are 
regular. 

Equation \eqref{route_equation} highlights the central role of the
pairing matrix $\Delta_{k \gamma l \delta}$ in determining the
symmetry properties of $F_{i\alpha j\beta}$. However, summation over
the intermediate states is cumbersome. Therefore, we convert Eq.
\eqref{route_equation} to k-space first:
\begin{equation}
F(\mathbf{k},\varepsilon)
= -G(\mathbf{k},\varepsilon) \Delta_{\mathbf{k}} G^{0*}(\mathbf{k},-\varepsilon)
  \label{k-route}
\end{equation}
where orbital indices are suppressed. Eq. \eqref{k-route} makes it clear that 
$F(\mathbf{k},\varepsilon)$ possesses the d-wave symmetry of 
$\Delta_{\mathbf{k}}$ since the regular Green's functions are rotationally 
invariant as shown in Appendix A.

Converting $F(\mathbf{k},\varepsilon)$ to the real-space, yields
\begin{equation}
  \label{TB-F}
  F_{i \alpha j \beta}(\varepsilon) = \sum_{\mathbf{k}} \langle i \alpha 
\vert \mathbf{k} \rangle
F(\mathbf{k},\varepsilon) \langle \mathbf{k} \vert j \beta \rangle,
\end{equation}
or 
\begin{equation}
  \label{TB-Fopen}
  F_{0 \alpha j \beta}(\varepsilon) =
  \sum_{\mathbf{k}}  e^{i(k_{z}z_{j}+k_{\parallel}\cdot R_{\parallel j})} \varphi^{*}_{\alpha}(\mathbf{k})\varphi_{\beta}(\mathbf{k})
  F(\mathbf{k},\varepsilon),
\end{equation}
where we have fixed the first site index to $R_i=0$, and made a separation
into perpendicular ($k_{z}z_{j}$) and parallel directions
($k_{x}x_{j}+k_{y}y_{j} = k_{\parallel} \cdot R_{\parallel j}$).

We first consider the case where orbitals $\alpha$ and $\beta$ are at the 
same horizontal site, i.e., $R_{\parallel j} = 0.$ The necessary condition 
for the matrix element $F_{0 \alpha j \beta}$ to be non-zero is that
\begin{equation}
  \label{onsite-f}
 \varphi^{*}_{\alpha}(\mathbf{k})\varphi_{\beta}(\mathbf{k}) F(\mathbf{k},\varepsilon),
\end{equation}
is rotationally invariant. Since $F(\mathbf{k},\varepsilon)$ is
d-wave, the product of wave functions on the right hand side of
Eq. \eqref{TB-Fopen} must have $d$-wave symmetry. For example, one of
the orbitals could be d-wave symmetric and the other rotationally
invariant. In particular, the anomalous matrix element between
$d_{x^2-y^2}$ and one of the set $\{s,p_{z},d_{z^2} \}$ satisfies this
condition.

Furthermore, if $z_{j}=0$, i.e., the orbitals are at the same site, the 
term \eqref{onsite-f} must also have an even parity in the z-direction for 
a non-zero matrix element, and hence $p_{z}$ would not be a possible pair 
with $d_{x^2-y^2}.$ However, in the case of $z_{j} \ne 0$, $p_{z}$ is 
allowed, since the phase factor $e^{ik_{z}z_{j}}$ does not have a 
well-defined parity. Note also that the anomalous matrix element between 
$d_{x^2-y^2}$ orbitals at the same horizontal position ($R_{\parallel 
j}=0$) is necessarily zero, since the term \eqref{onsite-f} is odd under 
rotations of $\pi/2$.

We next consider matrix elements between orbitals at neighboring sites 
where, $R_{\parallel j} = x_{j} = \pm a$, or $R_{\parallel j} = y_{j} = 
\pm a$. In this case, we will see that there will always be a non-zero 
matrix element with a properly symmetrized combination of neighboring wave 
functions, and the selection rules determine the correct choice of phase 
factors between sites. We discuss a particular case in detail as an 
exemplar. Specifically, let us compare sites $R_{j} = (a,0,c)$ and $R_{l} 
= (0,a,c)$ and check whether or not the sign of the sum in Eq. 
\eqref{TB-Fopen} changes. For the first site we get
\begin{equation}
 F_{0 \alpha j \beta} = \sum_{\mathbf{k}} e^{i(k_{x}a+k_{z}c)}
\varphi^{*}_{\alpha}(k_{x},k_{y})\varphi_{\beta}(k_{x},k_{y}) F(\mathbf{k},\varepsilon)
\end{equation}
 and for the second site
\begin{equation}
 F_{0 \alpha l \beta} = \sum_{\mathbf{k}} e^{i(k_{y}a+k_{z}c)}
\varphi^{*}_{\alpha}(k_{x},k_{y})\varphi_{\beta}(k_{x},k_{y}) F(\mathbf{k},\varepsilon).
\label{latter_eq}
\end{equation}
A rotation of $\pi/2$ is equivalent to the transformation $k_{y}
\rightarrow k_{x}$ and $k_{x} \rightarrow -k_{y}$.  Applying this  
to \eqref{latter_eq} yields
\begin{equation}
 F_{0 \alpha l \beta} = \sum_{\mathbf{k}} e^{i(k_{x}a+k_{z}c)}
\varphi^{*}_{\alpha}(-k_{y},k_{x})\varphi_{\beta}(-k_{y},k_{x}) F(\mathbf{k},\varepsilon).
\end{equation}

Thus the product 
$\varphi^{*}_{\alpha}(-k_{y},k_{x})\varphi_{\beta}(-k_{y},k_{x})$ 
determines what happens under rotations of $\pi/2$ around the site $i=0$.  
There are two cases: (1) This product is equal to 
$\varphi^{*}_{\alpha}(k_{x},k_{y})\varphi_{\beta}(k_{x},k_{y}),$ so that 
these terms are invariant, and the total effect of rotation on $F_{0\alpha 
j\beta}$ in Eq. \eqref{TB-Fopen} follows the d-wave symmetry of 
$F(\mathbf{k},\varepsilon)$; and (2) The products of the orbitals in 
k-space have opposite sign, and the matrix element $F_{0 \alpha j \beta}$ 
is invariant under in-plane rotation by $\pi/2.$ In either case there will 
be pairing between the central orbital $\alpha$ and a 
properly symmetrized orbital $\phi$, as defined in Eq. 
\eqref{eq:combination} of Appendix \ref{app:symmetry}. For case (1), an 
invariant $\varphi^{*}_{\alpha}(k_{x},k_{y})\varphi_{\beta}(k_{x},k_{y})$ 
linear combination of coefficients must be chosen with d-wave 
symmetry. For case 2, i.e., d-wave symmetric 
$\varphi^{*}_{\alpha}(k_{x},k_{y})\varphi_{\beta}(k_{x},k_{y})$, the 
correct linear combination has all positive expansion coefficients. 
Notably, for $\alpha = \beta$, the product of orbitals is invariant, so 
that any pair involving the same orbitals at neighboring sites must 
involve a linear combination of neighbors which is odd in rotations by 
$\pi/2.$ For example, in the anomalous matrix element $F_{78}$ between the 
$p_{z}$-orbitals of two Bi neighbors, the coefficient in the $x$-direction 
has an opposite sign to that in the $y$-direction.

\section{Discussion and conclusions}


We emphasize that the logic of symmetry rules for the anomalous matrix 
elements is more complicated than that of the regular matrix elements. In 
particular, the nonvanishing tunneling channels can be determined through 
group theoretic considerations\cite{NLMB, nieminenPRB}. For example, since 
the rotational symmetry of the $p_z$ orbitals of Bi and apical oxygen 
atoms differs from that of the $d_{x^2-y^2}$ orbital of the Cu {\it at the 
same horizontal position,} the corresponding off-diagonal term of the {\it 
regular} Green's function vanishes, inhibiting the corresponding tunneling 
channel. In contrast, coupling between electron and hole degrees of 
freedom via the gap matrix $\Delta_{\alpha \beta}$ leads to less obvious 
symmetry rules for the anomalous matrix elements: Now the quasiparticles 
are linear combinations of spin up electrons and spin down holes, and 
there is no simple rule for selecting the orbitals contributing to a 
chosen quasiparticle state. Hence, the $p_z$ or $d_{z^2}$ orbitals of Bi, 
O or Cu atoms may couple to a $d_{x^2-y^2}$ orbital of a Cu atom at the 
same horizontal position, and the possibility of this coupling must be 
checked by considering the tensor form of Dyson's equation, as written out 
in Eq. \eqref{k-route}, together with the transformation into 
tight-binding basis of Eq. \eqref{TB-Fopen}.


In summary, we have presented a comprehensive study of anomalous matrix 
elements of the Green's function derived from a realistic multiband model 
of Bi2212. The imaginary parts of these matrix elements describe the 
contributions of different orbitals to the coherence peaks involving the 
formation and breaking up of Cooper pairs. Although the pairing 
interaction is modeled by a local d-wave term in the Hamiltonian 
connecting only the $d_{x^2-y^2}$ orbitals of neighbouring Cu atoms, the 
anomalous matrix elements display a longer range with induced 
superconductivity appearing at other sites/orbitals, including the second 
cuprate layer and the BiO/SrO overlayers. Our analysis delineates the 
precise routes through which the induced superconductivity in a complex 
cuprate system is transferred between various orbitals and sites.

{\bf Acknowledgments} 
We acknowledge discussion with Matti Lindroos. This work is supported by 
the US Department of Energy contract DE-FG02-07ER46352 and benefited from 
the allocation of supercomputer time at NERSC and Northeastern 
University's Advanced Scientific Computation Center (ASCC). I.S. would 
like to thank Vilho, Yrj\"o ja Kalle V\"ais\"al\"a Foundation for 
financial support. This work benefited from resources of Institute of 
Advanced Computing, Tampere.

\appendix

\section{Symmetry properties of regular matrix elements of 
the Green's  function.}\label{app:symmetry}

This appendix delineates the symmetry properties of the regular matrix
elements of the normal and superconducting (SC) state Green's
functions $G^0(\epsilon)$ and $G(\epsilon)$, respectively, which were
seen in connection with Eq. \eqref{TB-Fopen} above to be important for
understanding the nature of anomalous matrix elements. Taking the
origin at the position of the $0th$ atom, $G^{0}_{0 \alpha j
  \beta}(\varepsilon)$ can be written as
\begin{equation}
  \label{eq:Grealreciprocal}
\begin{split}
  G^{0}_{0 \alpha j \beta}(\varepsilon)& = \sum_{\mathbf{k}} \langle 0 \alpha \vert
  \mathbf{k} \rangle
  G^{0}(\mathbf{k},\varepsilon) \langle \mathbf{k} \vert j \beta \rangle\\
  & = \sum_{k} e^{i\mathbf{k}\cdot \mathbf{R_{j}}}
  \varphi^{*}_{\alpha}(\mathbf{k})\varphi_{\beta}(\mathbf{k})
  G^{0}(\mathbf{k},\varepsilon),
\end{split}
\end{equation}
where
\begin{equation}
\label{geekoo}
  G^{0}(\mathbf{k},\varepsilon) = \frac{1}{\varepsilon-\varepsilon_{k}-\Sigma(\varepsilon)}.
\end{equation}
Since the Hamiltonian is invariant under rotations of $\pi/2$, the 
dispersion $\varepsilon_{k}$ and $G^{0}(\mathbf{k},\varepsilon)$ are also 
invariant.  This is true as well for the SC state regular Green's function 
since the self-energy in Eq. \eqref{geekoo} is augmented by an additional 
term $\Sigma^{BCS} = \Delta_{k} G^{0}_{h}(\mathbf{k},\varepsilon) 
\Delta^{\dagger}_{k},$ which is rotationally invariant\cite{nieminenPRB}. 
Because $G^0$ and $G$ possess the same symmetry properties, in the following, 
we only consider the symmetry properties of $G^{0}(\mathbf{k},\varepsilon)$.

Consider first the case where $R_{j} = (0,0,c).$ Then, $G^{0}_{0 \alpha j
  \beta} \ne 0$ only if
$\varphi^{*}_{\alpha}(\mathbf{k})\varphi_{\beta}(\mathbf{k})$ in Eq.
\eqref{eq:Grealreciprocal} is invariant under the in-plane operations
of the symmetry group of the Hamiltonian. For example, a $p_z$ orbital
can have non-zero matrix elements with $s$, $p_z$ or $d_{z^2}$ of an
atom at the same horizontal position. But the matrix element between
$p_z$ and $d_{x^2-y^2}$ of atoms at the same horizontal position is
zero. For $c=0$, the matrix element is non-zero only if the orbitals
are similar.

We next consider the case where there are four atoms around a central atom 
at the distance of the horizontal lattice constant $a:$ $R_{1}=(a,0,c),$ 
$R_{2}=(0,a,c),$ $R_{3}=(-a,0,c),$ and $R_{4}=(0,-a,c).$ Changing the 
indices according to $1 \rightarrow 2 \rightarrow 3 \rightarrow 4 
\rightarrow 1$ corresponds to rotations by $\pi/2$ in real space. 
The transformation $k_{y} \rightarrow k_{x}$ and $k_{x} \rightarrow 
-k_{y}$ 
represents the same rotation in k-space. Now the phase factor $e^{ik\cdot 
R_{j}}$ has a fundamental effect on the symmetry behavior. Let us compare 
cases $R_{1} = (a,0,c)$ and $R_{2} = (0,a,c)$ and check whether the sign 
of the sum changes. In the first instance we get
\begin{equation}
 G^{0}_{0 \alpha 1 \beta}(\varepsilon) = \sum_{\mathbf{k}} e^{i(k_{x}a+k_{z}c)}
\varphi^{*}_{\alpha}(k_{x},k_{y})\varphi_{\beta}(k_{x},k_{y}) G^{0}(\mathbf{k},\varepsilon)
\end{equation}
while the second case gives
\begin{equation}
 G^{0}_{0 \alpha 2 \beta}(\varepsilon) = \sum_{\mathbf{k}} e^{i(k_{y}a+k_{z}c)}
\varphi^{*}_{\alpha}(k_{x},k_{y})\varphi_{\beta}(k_{x},k_{y}) G^{0}(\mathbf{k},\varepsilon)
\label{latter_app_eq}
\end{equation}
Applying the transformation $k_{y} \rightarrow k_{x}$ and
$k_{x} \rightarrow -k_{y}$ to \eqref{latter_app_eq} yields
\begin{equation}
 G^{0}_{0 \alpha 2 \beta}(\varepsilon) = \sum_{\mathbf{k}} e^{i(k_{x}a+k_{z}c)}
\varphi^{*}_{\alpha}(-k_{y},k_{x})\varphi_{\beta}(-k_{y},k_{x}) G^{0}(\mathbf{k},\varepsilon).
\end{equation}

Thus, it is the product 
$\varphi^{*}_{\alpha}(-k_{y},k_{x})\varphi_{\beta}(-k_{y},k_{x})$ that 
determines what happens under rotations of $\pi/2.$ If this is equal to 
$\varphi^{*}_{\alpha}(k_{x},k_{y})\varphi_{\beta}(k_{x},k_{y}),$ the 
matrix element does not change sign under rotations, otherwise $G^{0}_{0 
\alpha j \beta}$ changes sign under in-plane rotation of $\pi/2.$ In 
particular, for $\alpha=\beta$, this term is invariant, but for $\alpha = 
p_{z}$ or $d_{z^2}$ and $\beta = d_{x^2-y^2},$ there is a change of sign 
under rotation.

An equivalent approach is to consider a linear combination of orbitals
$j \beta$
\begin{equation}
  \label{eq:combination}
  \vert \phi \rangle = \sum_{j=1}^{4} c_{j} \vert j \beta \rangle.
\end{equation}
There is a non-zero regular matrix element $G^{0}_{0\alpha, \phi}$ only if
$\vert \phi \rangle$ belongs to the same representation of the
symmetry group of the Hamiltonian as orbital $\vert 0 \alpha \rangle.$
The transformation of the expansion coefficients $c_{j}$ directly
follows from the transformation of $G^{0}_{0 \alpha j \beta}.$
For example, it is obvious that 
\begin{equation}
  \label{eq:combigreen}
  G^{0}_{0 \alpha, \phi} = \sum_{j=1}^{4}c_{j} G^{0}_{0 \alpha j \beta},
\end{equation}
Hence, if $\alpha = \beta = d_{x^2-y^2}$ and $j=1...4$ are defined as
above, $c_{j}$ must be a constant in order to keep the full symmetry of
the group of the Hamiltonian, leading to $$G^{0}_{0 \alpha, \phi}
\propto
c_{1}e^{ik_{z}c}[\cos{(k_{x}a)}+\cos{(k_{y}a)}](k_{x}^2-k_{y}^2)^2$$
If, however, $\alpha = d_{z^2}$ and $\beta = d_{x^2-y^2}$, one must
have $c_{2} = c_{4} = -c_{1} = -c_{3}$, and then $$G^{0}_{0 \alpha, \phi}
\propto 
c_{1}e^{ik_{z}c}[\cos{(k_{x}a)}-\cos{(k_{y}a)}](k_{x}^2-k_{y}^2)(3k_{z}^2-k^2),$$ 
which requires change of sign of $c_{j}$'s under rotations of $\pi/2$ 
in order to obtain an invariant matrix element.


\begin{thebibliography}{99}
\bibitem{ABfoot1}Our use of a relatively thin slab of seven layers is
  adequate for our purpose of exploring tunneling to the tip from the
  cuprate bi-layer in the presence of BiO and SrO overlayers. Thicker
  slabs containing multiple unit cells will be needed for delineating
  c-axis coherence effects involving coupling between different
  cuprate bi-layers.\cite{Spathis} Also, since our modeling involves
  BiO and SrO layers with the perfect bulk structure, it is not
  sensitive to effects of impurities in the surface layers. O, Ni and
  Zn impurities have been shown to produce a variety of interesting
  features in the STS spectrum.\cite{Hudson,Pan,Hirschfeld}.
\bibitem{ABfoot4}
The Hamiltonian of Eq. (1) is appropriate for the overdoped regime 
considered here. In the underdoped system, effects of the pseudogap must 
of course also be included as, for example, in Refs. \onlinecite{Das,Basak}.
\bibitem{RMfoot1}Within the LDA, there is a weak hybridization between
  the Cu-$d$ and Bi-$p_z$ orbitals, which plays a significant role in
  producing the observed tunnel current.
\bibitem{ABfoot2}Doping independent tight-binding parameters in the
  spirit of a rigid band picture are implicit in the form of our
  Hamiltonian in Eq. (1). A more realistic treatment of doping effects
  on electronic states could be undertaken using various approaches
  (see, e.g., Refs.~\onlinecite{AB1,AB2,AB3}). However, we expect the
  rigid band model to be a good approximation for doping away from the
  Cu-O planes.
\bibitem{ABfoot3}We have used a sparse $\Delta$ matrix limited to
  coupling between $d_{x^{2}-y^{2}}$ orbitals of neighboring Cu atoms
  only in order to make the relation between $\Delta$ and $\langle
  c_{\alpha \downarrow}c_{\beta \uparrow} \rangle$ tractable. A more
  general $\Delta$ matrix could be obtained via a self-consistent
  application of Eq. \eqref{eq:gapgen}.
\bibitem{RMfoot2}Note that in order to properly describe the shape of the
  coherence peaks\cite{Fang} we would need to include explicit
  coupling to bosonic modes.
\bibitem{ABfoot5}For our illustrative purposes, we have taken the
  imaginary part of the self-energy in the overdoped regime to be of
  Fermi liquid form\cite{nieminenPRB}. A more realistic 
description would consider
  self-energy corrections to account for
  electron-electron\cite{Das} and electron-phonon\cite{elphon}
  interaction effects.
\bibitem{rotationfoot} By `rotationally invariant' we mean invariance
  to in-plane rotation only, e.g., $d_{z^2}$ and $p_z$ orbitals are
  rotationally invariant.  Also, by d-wave symmetry we mean an odd
  parity with respect to rotations of $\pi/2$ around the $z$-axis.
\end{thebibliography}
\end{document}